\documentclass[prb,twocolumn,amsmath,amssymb,superscriptaddress,showpacs]{revtex4-1}
\usepackage{graphicx}
\usepackage{graphicx}
\usepackage{dcolumn}
\usepackage{bm}
\usepackage{mathrsfs}
\usepackage{subfigure}
\usepackage{natbib}
\usepackage{color}
\usepackage{amssymb}
\usepackage{amsmath}
\usepackage{cancel}
\usepackage{bbm}
\usepackage{float}
\usepackage[breaklinks,colorlinks,citecolor=blue]{hyperref} 

\usepackage{changes}
    \colorlet{Changes@Color}{blue}

\begin{document}

\title{Vortex Clusters in Thin Superconductors: Extended Ginzburg-Landau Analysis}

\title{Vortex interaction in thin films - a crossover from type I to type II superconductivity}

\author{W.  Y. C\'{o}rdoba-Camacho}
\affiliation{National Research University Higher School of Economics, 101000 Moscow, Russia}

\author{A. Vagov}
\affiliation{Institute for Theoretical Physics III, University of Bayreuth, Bayreuth 95440, Germany}
\affiliation{ITMO University, St. Petersburg, 197101, Russia}

\author{A. A. Shanenko}
\affiliation{Departamento de F\'isica, Universidade Federal de Pernambuco, Av. Jorn. An\'ibal Fernandes, s/n, Cidade Universit\'aria 50740-560, Recife, PE, Brazil}

\author{J. Albino Aguiar}
\affiliation{Departamento de F\'isica, Universidade Federal de Pernambuco, Av. Jorn. An\'ibal Fernandes, s/n, Cidade Universit\'aria 50740-560, Recife, PE, Brazil}

\author{V. S. Stolyarov}
\affiliation{Moscow Institute of Physics and Technology, 141700 Dolgoprudny, Russia}
\affiliation{Dukhov Research Institute of Automatics (VNIIA), 127055 Moscow, Russia}

\author{A. S. Vasenko }
\affiliation{National Research University Higher School of Economics, 101000 Moscow, Russia}
\affiliation{I.E. Tamm Department of Theoretical Physics, P.N. Lebedev Physical Institute, Russian Academy of Sciences, 119991 Moscow, Russia}

\date{today}

\begin{abstract} 

Interactions between vortices in thin superconducting films are investigated in the crossover (intertype) regime between superconductivity types I and II. We consider two main factors responsible for this crossover: a) changes in the material characteristics of the film and b) variations of the film thickness controlling the effect of the stray magnetic fields outside superconducting sample. The analysis is done within the formalism that combines the perturbation expansion of the microscopic equations to one order beyond the Ginzburg-Landau theory with the leading contribution of the stray fields. It is shown that the latter gives rise to qualitatively different spatial profile and temperature dependence of the vortex interaction potential, as compared to bulk vortex interactions. The resulting interaction is long-range repulsive while exhibiting complex competition of attraction and repulsion at small and intermediate separations of vortices. This explains the appearance of vortex chains reported earlier for superconducting films. 
\end{abstract}

\pacs{74.20.De,74.25.Dw,74.25.Ha,74.78.-w}
\maketitle
\section{Introduction.}
Magnetic properties of superconductors are routinely classified as type I or type II by the value of the Ginzburg-Landau (GL) parameter $\kappa = \lambda_L/\xi$ ($\lambda_L$ is the London magnetic penetration depth and $\xi$ is the GL coherence length). It is also well known that when $\kappa$ of a superconductor is close to the critical value $\kappa_0 = 1/\sqrt{2}$, its magnetic response is neither of the two standard types. Such materials are in the crossover or the intertype (IT) regime and demonstrate the intermediate mixed state (IMS), where the magnetic flux penetrates the superconducting condensate in the forms of unconventional patterns such as vortex clusters (islands) embedded in the Meissner state.\cite{krag,essmann,jacobs, auer,lav1,lav2, muhlb1,brandt1,pau,ge1, muhlb2,muhlb3,ge2} 

Earlier works related the IMS with the non-standard vortex patterns in IT superconductors to the fact that in this regime the vortex-vortex interactions are repulsive at small but attractive at large distances. This {\it type II/1} concept was extended recently\cite{vag1,wolf1} by elaborating its relation with the critical Bogomolnyi (B) point\cite{wein, bogomol1, bogomol2}, at which the condensate state is self dual and infinitely degenerate. It was shown that the IT superconductivity is closely connected to mechanisms that remove this degeneracy which gives rise,  in particular, to the non-monotonic vortex-vortex interaction. Other consequences include the enhancement of the many-vortex interactions that stabilize large vortex clusters even when the pairwise interaction is fully attractive.\cite{wolf2}

\begin{figure}[]
	\begin{center}
		\includegraphics[width=0.3\textwidth]{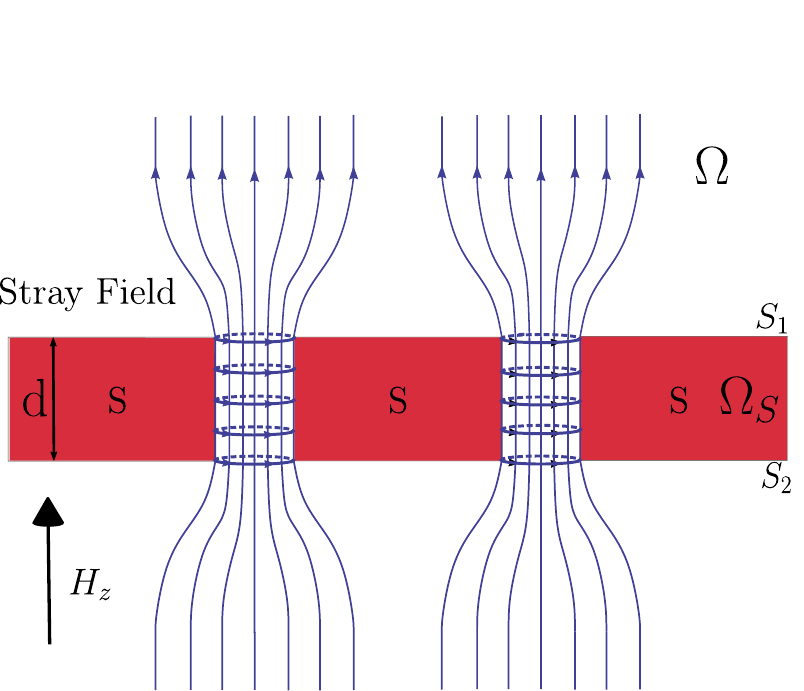}
	\end{center}
	\caption{A sketch of the system: superconducting (S) film with thickness $d$ placed in a perpendicular magnetic field ${\bf H}$ directed along the $z$-axis and penetrated by magnetic vortices. }
	\label{fig82}
\end{figure}

The degeneracy can be removed by many physically different mechanisms. One of these is controlled by the material properties that define the value of $\kappa$ and  non-local interactions within the condensate state at $T<T_c$, creating the IT domain in the $\kappa - T$ plane for bulk samples\cite{vag1} which, among other things, is characterized by the non-monotonic vortex interactions.\cite{wolf1} In finite and low-dimensional samples the degeneracy can also be removed by the interactions with the stray magnetic field outside the sample and its boundaries. Indeed, it is long known that the stray field makes a type I superconducting film to become a type II superconductor when its thickness decreases.\cite{tinkh,maki} As in the bulk case, in a film the interchange between the types is not instantaneous, occupying a finite thickness interval.  In the crossover regime the film mixed state demonstrates many exotic spatial configurations, some of which are similar to bulk IT superconductors.\cite{silcox, has,cor1}  The crossover interval with unconventional vortex patterns was also predicted for superconducting wires.\cite{cor2} It is not yet clear, however, if and how those geometry related factors, which lift the BP degeneracy, affect the interactions between vortices.

In this work we investigate details of how vortex interactions are changed in the crossover regime in a thin film, where the superconductivity type is altered by a combination of the stray fields input and the material related factors, that give rise to the bulk IT superconductivity. This analysis is done by extending the perturbation expansion with respect to proximity to the critical temperature $\tau=1-T/T_c$, developed earlier for the microscopic equations in bulk superconductors  \cite{shan,vag2,vag3}, to the case of films with the contribution of the stray fields.

\section{Vortex interaction energy}

\subsection{Model and assumptions}

The calculations are done for a geometry schematically shown in Fig.~\ref{fig82}: a superconducting film of thickness $d$ is placed in the external magnetic field ${\bf H} = (0,0,H)$ perpendicular to the film surface. The vortex interaction potential is obtained from the Gibbs free energy
 \begin{align}
 G=F+ \int d^3 {\bf r} \left( \frac{\boldsymbol{B}^2}{8\pi} - \frac{\boldsymbol{B}H}{4\pi} \right),
 \label{eq:Gibbs}
  \end{align}
calculated for a given vortex configuration, where $F$ is the corresponding condensate free energy and ${\bf B}$ is the magnetic field. The latter satisfies the asymptotic condition ${\bf B} \to {\bf H}$ at infinity, ensuring that the magnetic flux is the same for both ${\bf B}$ and ${\bf H}$.

The Gibbs free energy is calculated by taking into account the superconducting condensate and the magnetic flux both inside and outside the superconductor. The calculation is split in two parts. First, the superconducting state is found inside the sample, assuming that it is the same as in the bulk system. Second, the stray field is calculated from the appropriate boundary condition at the film surface and at infinity. The approach is valid when the stray field has a relatively weak reciprocal effect on the superconducting state inside the sample. This assumption strictly holds when the film is not ultrathin. It should also be noted that describing IT effects in bulk samples requires the approach beyond the GL theory, in contrast to the stray field contributions that can be taken into account within the GL approach.  

%

\begin{figure}[]
	\begin{center}
		\includegraphics[width=0.35\textwidth]{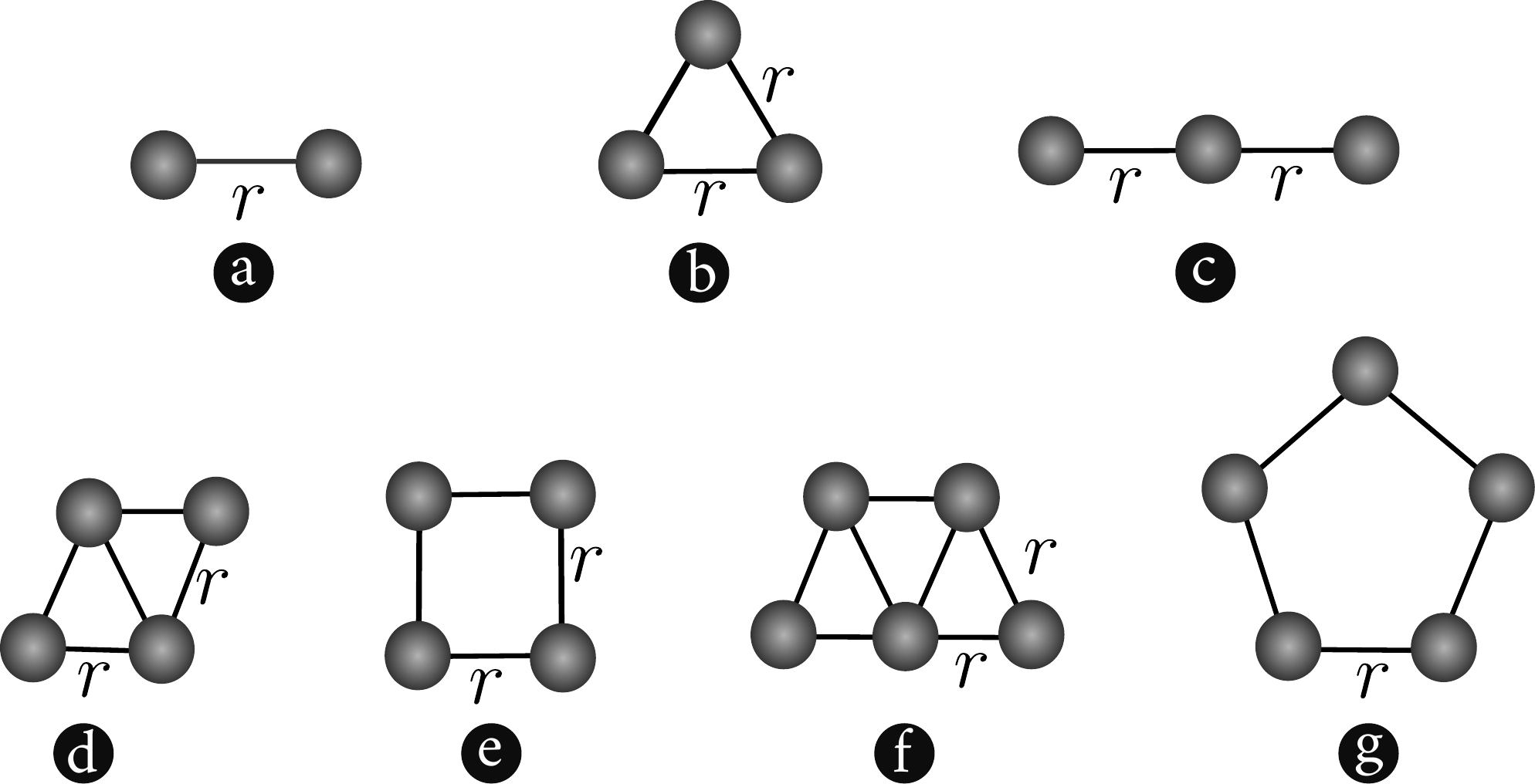}
	\end{center}
	\caption{Clusters of equidistant vortices used to calculate vortex interaction potential as a function of the inter-vortex distance $r$.}
\label{fig83}
\end{figure}  

\subsection{Energy of the superconducting film}

The Gibbs free energy of a superconducting film $G_s$ is obtained within the extended GL (EGL) formalism derived as a perturbative expansion of the microscopic BCS equations with respect to $\tau$, where only the leading corrections to the GL theory are kept. Since our work is focused on the IT regime in the vicinity of the B point $(\kappa_0, T_c)$, we also apply the perturbation expansion with respect to the deviation $\delta \kappa = \kappa - \kappa_0$ from the critical GL parameter. 

This expansion was derived earlier and here we only quote the final result. However, due to the geometry of the problem the result has to be modified slightly. Earlier works calculated the difference ${\cal G}_s$ between the Gibbs free energies of the nonuniform and uniform (Meissner) states at the thermodynamic critical field $H=H_c$, see Ref.~\onlinecite{vag1}. In contrast, here the magnetic field $H$ is equal to the average magnetic field inside the film. Taking this into account, we express the Gibbs free energy as 
\begin{equation}
\label{Eq4}
G_s = {\cal G}_s  - \int_{\Omega_S} d^3{\bf r}  \left( \frac{H_c^2}{8\pi} + \frac{\boldsymbol{B}\delta H}{4\pi} \right),
\end{equation}
where $\delta H = H - H_c$ and the integral is taken over the volume $\Omega_S$ of the sample. The integral in Eq.~(\ref{Eq4}) depends only on the number of vortices but not on their positions and can be omitted in the analysis of the vortex interactions. Thus, we obtain the two leading contributions to the Gibbs free energy, that affect the vortex interaction potential, as 
\begin{align}
G_{s} = &G_0 \,  \frac{\tau\, d}{\xi(0)} \left\{  - \sqrt{2}\,  {\cal I} \, \delta \kappa
+ \Big[ C_1 \,{\cal I} + C_2 {\cal J} \Big] \,  \tau \right\}, 
\label{eq:functional_exp_final}
\end{align}
where the GL contribution at $\kappa=\kappa_0$ vanishes, reflecting the degeneracy of the B point and the energy unit is $G_0 = H_{c}^2(0) \xi^3(0)/4\pi$, with $H_c(0)$ and $\xi(0)$ the thermodynamic critical field and the coherence length of the GL theory taken at $T=0$. Dimensionless constants $C_1$ and $C_2$ depend on the microscopic model for the carrier bands; for the single-band case with a spherical Fermi surface one obtains universal constants $C_1\approx -0.41$ and $C_2 \approx 0.68$, which do not depend on microscopic parameters such as the Fermi velocity or the density of states at the Fermi surface. Two other constants in Eq.~(\ref{eq:functional_exp_final}) are given by 2D integrals taken over the $x-y$ plane 
\begin{align}
{\cal I} =\!\! \int |\Psi|^2 \big(1 - |\Psi|^2\big)d^2{\bf r},\;{\cal
J} =\!\! \int |\Psi|^4 \big(1 - |\Psi|^2\big)d^2{\bf r},
\label{eq:I_J}
\end{align}
and depend only on the order parameter $\Psi$ which is found by solving the dimensionless GL equations taken at $\kappa_0$ and given by
\begin{align}
\Psi -\Psi|\Psi|^2 + {\bf D}^2\Psi=0,\; \boldsymbol{\nabla}\times {\bf B}=2{\rm Im}\left[\Psi{\bf D}^*\Psi^*\right],
\end{align}
where ${\bf D}=\boldsymbol{\nabla}+\mathbbm{i}{\bf A}$ and all the pertinent quantities are scaled as ${\bf B}\to {\bf B}/H_c(0)$, ${\bf A}\to {\bf A}/\xi(0)H_c(0)$, ${\bf r} \to {\bf r}/\xi(0)$, and $\Psi \to \Psi/\Psi_0$, with $\Psi_0$ the uniform solution to the GL equations.  

At $\kappa_0$ the GL equations have a self-dual form, where the dimensionless magnetic field and the order parameter are related algebraically as $B = 1 - |\Psi|^2$.\cite{bogomol1,bogomol2} If the order parameter is sought in the form $\Psi = e^{- \varphi} \psi$, the GL equations are written as 
\begin{align} 
& \big(\partial_x^2 +  \partial_y^2\big) \varphi  = 1 - e^{-2  \varphi } |\psi|^2, \quad (\partial_x + i \partial_y ) \psi = 0.
\label{eq:self-dual}
\end{align}
A solution to these equations can be obtained for a configuration with arbitrary vortex number and positions by noting that the right equation in Eqs.~(\ref{eq:self-dual}) is satisfied by any analytical function of the complex variable $z=x+ \mathbbm{i} y$. The solution with $N$ vortices located at the points $z_i = (x_i,y_i)$ is obtained as $\prod_{i=1}^N (z - z_i)$, which is then substituted into Eq.~(\ref{eq:self-dual}). In practical calculations we seek the $N$-vortex solution as a product 
\begin{align}
\Psi (z)= e^{-\delta \varphi} \prod_{i=1}^N  \Psi_i( z), \quad  \Psi_i( z) = \Psi^{(1)}( z - z_i),
\label{eq:subst}
\end{align}
where $\Psi^{(1)}(z)$ is the solution for a single vortex with the center at $z=0$. Function $\delta \varphi$ is then found from the equation
\begin{align}
\big(\partial_x^2 +  \partial_y^2\big)\delta\varphi = &\sum \limits_i
\big(|\Psi_i|^2 - 1) +1-\prod \limits_i\big|\Psi_i\big|^2e^{-2\delta \varphi},
\label{eq:delta_phi}
\end{align} 
which is solved numerically by assuming the asymptotic condition $\delta \varphi \to 0$ far from the vortex cores. 

\subsection{Energy of the stray field}

The stray field contribution to the Gibbs free energy $G_{f}$ is calculated using Eq.~(\ref{Eq4}), where $F = 0$ and the integration is taken over the space outside the sample. In order to calculate the part of the stray field energy due to the presence of a superconducting film, we subtract the energy of the homogeneous external field ${\bf H}^2/8\pi$. The field ${\bf B}$ is found by solving the magnetostatic problem $\boldsymbol{\nabla}\times\boldsymbol{B} = 0$ and $\boldsymbol{\nabla} \cdot\boldsymbol{B} =0$ with the boundary conditions defined at the film surface and at infinity. Following the earlier assumption that the stray field does not affect the field inside the superconductor, the field at the boundary of the sample is given by the solution of the GL equations. 

The magnetostatic problem is solved by following a standard routine and introducing the scalar potential $\phi$ for the magnetic field, so that $\boldsymbol{B}-{\bf H} = -\boldsymbol{\nabla}\phi$. With this choice the first equation of the problem is satisfied identically while the second gives the Laplace equation
\begin{equation}
\label{Eq13}
    \boldsymbol{\nabla}^2\phi=0, 
\end{equation}
where we assume $\phi \to 0$ far from the film, $z \to \pm \infty$. The solution must satisfy the boundary conditions at the film surface
\begin{align}
\label{Eq14}
\big({\bf n} \cdot \boldsymbol{\nabla}\big) \phi = H - B(\boldsymbol{\rho}),
\end{align}
where ${\bf n}$ denotes the unit vector normal to the surface $S$, $\boldsymbol{\rho} = (x,y)$ are the in-plane surface coordinates, and $B(\boldsymbol{\rho})$ is the magnetic field at the boundary obtained from the GL equations inside the film. The magnetostatic problem is solved separately above and below the sample with the only difference that the in-plane component of the field changes its sign, so that $\phi(\boldsymbol{\rho}, z) = - \phi ( \boldsymbol{\rho}, - z)$, assuming the center of the film at $z=0$.

The solution to the Laplace equation above the film writes as
\begin{align}
\label{Eq16}
\phi(\boldsymbol{r})=-\frac{1}{2\pi}\int_{S} \frac{B (\boldsymbol{\rho})}{|\boldsymbol{r}-\boldsymbol{\rho}|}d^2\boldsymbol{\rho},
\end{align}
where the integral is taken over the upper film surface $S$. The magnetic field is obtained by taking the potential gradient as
\begin{align}
\label{Eq17}
\boldsymbol{B}(\boldsymbol{r}) - \boldsymbol{H}=\frac{1}{2\pi}\int_{S}  \big(B(\boldsymbol{\rho}) - H\big) \nabla_{\bf r} \,\frac{1}{|\boldsymbol{r} -\boldsymbol{\rho}|}d^2\boldsymbol{\rho}.
\end{align}
The corresponding energy is calculated by substituting this expression into Eq.~(\ref{eq:Gibbs}).  After some manipulations one obtains the stray field contribution to the energy as
\begin{equation}
\label{Eq21}
G_{f} =\frac{G_0}{2\pi}\sqrt{\tau}\int_{S} \int_{S} \frac{\big(B (\boldsymbol{\rho})-H\big) \big(B (\boldsymbol{\rho}^\prime)-H\big)}{|\boldsymbol{\rho}-\boldsymbol{\rho}^\prime|} d^2\boldsymbol{\rho}^\prime d^2\boldsymbol{\rho},
\end{equation}
where we use the dimensionless quantities and take into account that the contributions above and below the film are equal.

We note that the approximations adopted to calculate the contribution of the stray field assumes that inside the film the parallel component of the field can be neglected, so the vortex width does not change with $z$. However, as follows from Eq. (\ref{Eq16}) parallel component ${B}_\parallel$ is non-zero outside the film. It describes vortex widening outside the film and contributes to the energy. We also note that this calculation of the stray field contribution is done within the GL theory since it uses the GL solution as the boundary condition. This approximation is appropriate when the value of the stray field contribution is similar to the leading contributions to the Gibbs energy in  Eq. (\ref{eq:functional_exp_final}), which may break for very thin films.  We note, however, that in practice our approximations lead to an underestimation of the stray filed contribution but does not result in qualitative changes. One can therefore formally extend the results to the limit of very thin films.

\begin{figure*}
	\begin{center}
		\includegraphics[width=0.7 \textwidth]{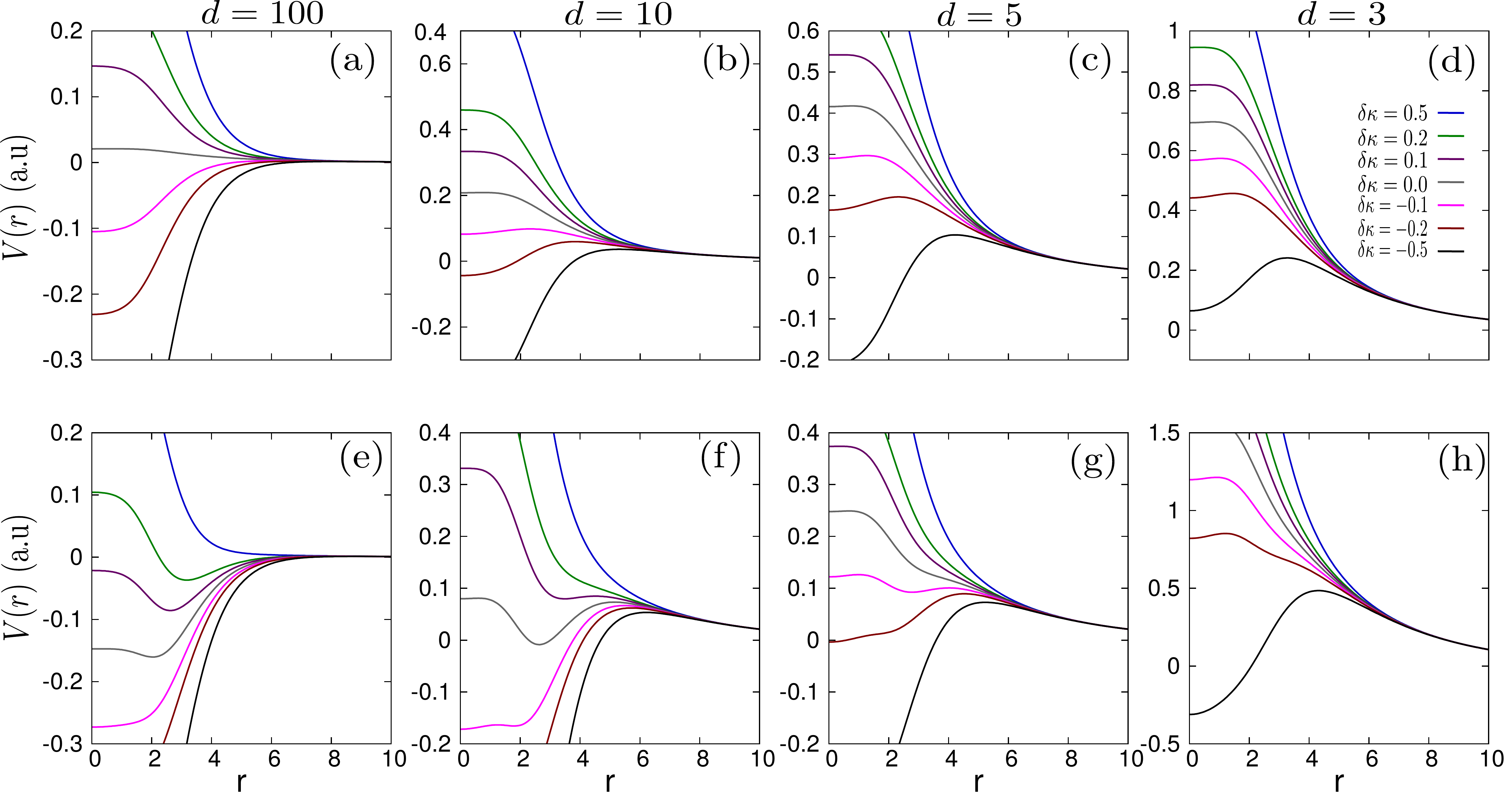}
	\end{center}
	\caption{Pair vortex potential, calculated within the GL theory [upper panels (a-d)] and by using the EGL formalism [lower panels (e-h)], for the film thicknesses $d/\xi(0)=100, 10, 5, 3$. Each panel shows the interaction potential calculated for $\delta\kappa = 0.5, 0.2, 0.1, 0.0, -0.1, -0.2$ and $-0.5$.}
	\label{Fig31}
\end{figure*}

\section{Vortex interactions}

The potential energy of the vortex interaction is given by the sum of the position-dependent Gibbs free energies inside and outside the sample, from which one should subtract the energy of $N$ isolated vortices $N(G_{s}^{(1)}+ G_{f}^{(1)})$. The resulting interaction potential per vortex is given by
\begin{align}
V = \big(G_{s} + G_{f}\big)/N - G_{s}^{(1)} - G_{f}^{(1)}.
\label{int}
\end{align}
Using the developed formalism we can calculate the interaction potential for an arbitrary vortex configuration by obtaining the corresponding solution to the GL equations (\ref{eq:self-dual}) and then substituting this solution into Eqs.~(\ref{eq:functional_exp_final}), (\ref{Eq21}), and (\ref{int}).  Before proceeding to numerical calculations, we make several general observations that follow from the obtained energy expressions.

\subsection{General properties}

First we note that following the obtained expressions quantitative details of the vortex interactions are qualitatively independent of the material parameters. It is clear that the stray field leads to an additional vortex repulsion. It is also easy to see that the film contribution to the interaction does not depend much on its material parameters. Indeed, according to Eq.  (\ref{eq:functional_exp_final}) the latter affect the interaction potential via  constants $C_{1,2}$ and the GL parameter $\kappa$. It can be demonstrated that the structure of the IT domain remains qualitatively the same as long as $C_2 >0$, which ensures that the IT domain has a non-zero width.\cite{vag1} The vortex interaction is given by the interplay between the stray field and the bulk IT factors, with the relative contributions depending on the ratio between those few parameters, which can change the quantitative dimensions of the crossover interval but not its qualitative characteristics. 

In particular, one notes that the bulk and the stray field contributions to Eq. (\ref{eq:functional_exp_final}) have different temperature dependencies. The stray field contribution is $\propto \sqrt{\tau}$ becoming dominant in the limit $\tau \to 0$. This contribution is also dominant in the limit of small widths $d \to 0$. This confirms the known conclusion that a superconductive film is always a type II superconductor when it is very thin or is close to the critical temperature. 

Finally, the asymptotic behaviour of the stray field contribution at large inter-vortex distances $r$ differs from what one observes in bulk superconductors. In a bulk sample the vortex interaction potential in Eq.  (\ref{eq:functional_exp_final}) is determined by functions ${\cal I}$ and ${\cal J}$, which decay exponentially $\exp(-r)$ at large inter-vortex distance $r$. In contrast, the stray-field contribution has the power law asymptotic $\propto 1/r$, which dominates at large distances leading to the fact that the long-range vortex interaction is always repulsive. Notice that this asymptotic law agrees with the long-range interaction between the Pearl vortices in ultra-thin films\cite{brandt2} which illustrates the earlier remark that although our results are derived under the assumption that the film is not very thin, they can be formally extended to the limit $d\to 0$, giving qualitatively correct results. 

\begin{figure*}
\begin{center}
\resizebox{1.99\columnwidth}{!}{\rotatebox{0}{
\includegraphics{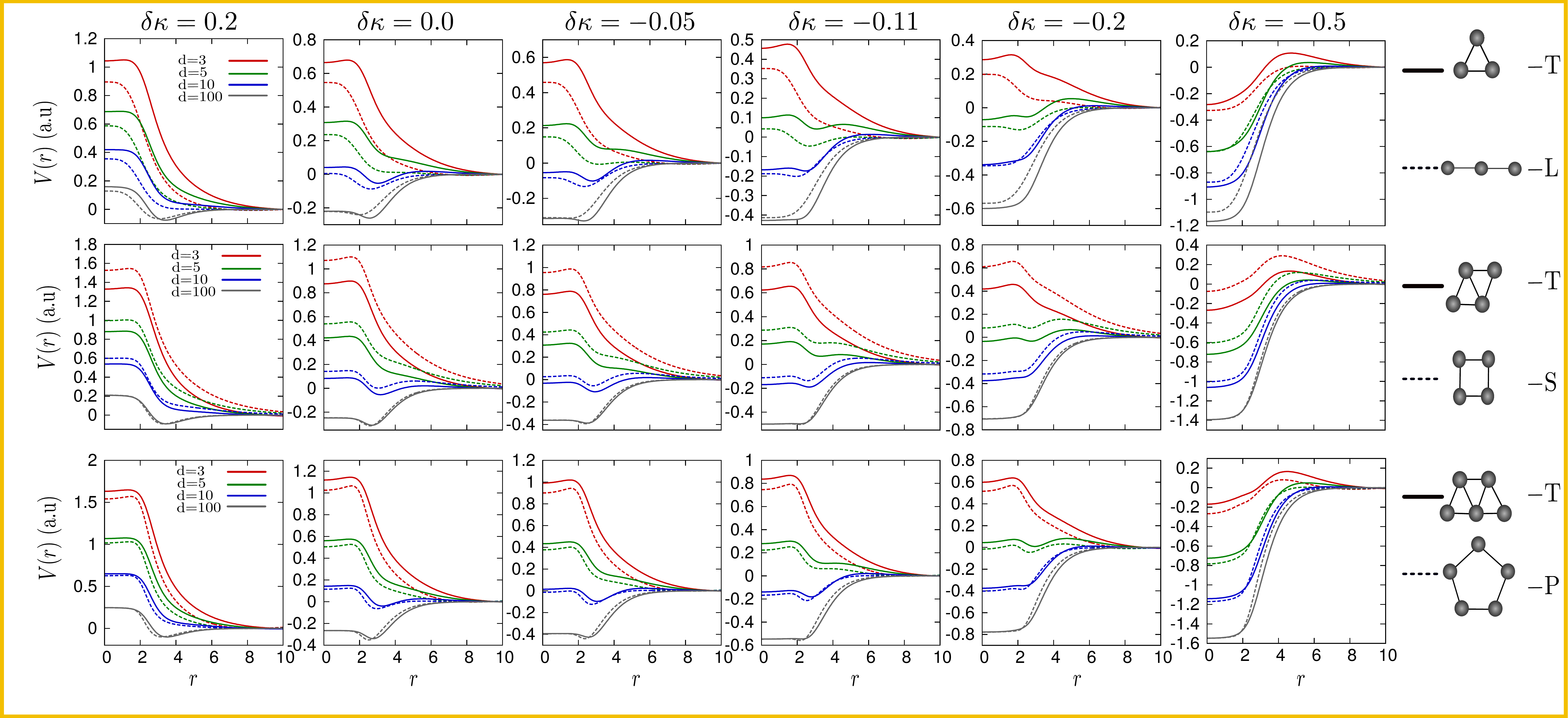}}}
\end{center}
 	\caption{Interaction potential for different vortex clusters (shown on the right)  calculated for films with the thickness $d/\xi(0) = 2,5,10,100$ for $\delta \kappa = -0.5, -0.2, -0.11, -0.05, 0, 0.2$.}
\label{fig88}
\end{figure*} 

\subsection{Pair vortex interactions}

Before studying the general case of vortex interactions in many-vortex clusters we consider the simpler case of two vortices, comparing the GL and EGL results. The pairwise potential is shown in Fig. \ref{Fig31}, where the upper panels give the calculations with only GL contributions for the superconducting sample and the lower panels show the complete EGL result. The columns from right to left correspond to the widths $d=100, 10, 5, 3$ in the units of $\xi(0)$, respectively. The lines represent materials with several values of $\delta \kappa$ taken in the interval $[-0.5,0.2]$.  

The case $d=100$ is essentially the bulk limit with a negligible intrusion by the stray field. Consequently, results in Figs. \ref{Fig31} a) and e) allow one to estimate the difference between the GL and EGL approaches for the bulk sample. As expected, according to the GL theory [Fig. \ref{Fig31} a)] the IT domain shrinks into a single point $\delta \kappa =0$, so that at $\delta \kappa >0 $ the vortex-vortex interaction is repulsive (type II behaviour) while at $\delta \kappa < 0 $ it is attractive (type I). In contrast, the EGL approach yields a finite IT domain [Fig. \ref{Fig31} e)] where the pair vortex interaction potential is spatially non-monotonic - attractive at large and repulsive at a small distance $r$.  

In the opposite limit of very thin films $d=3$, the stray fields dominates the interaction and the difference between the GL and EGL becomes irrelevant [Figs. \ref{Fig31} d) and h)]. One observes either a monotonic repulsion (for $-0.2 < \delta \kappa $) or a non-monotonic dependence for the deep type I materials (for $\delta \kappa < -0.5$).    

One notes that, in contrast to bulk IT superconductors, where the interaction is repulsive at small and attractive at larger distances [Fig. \ref{Fig31} e)], here the interaction is repulsive at large distances but attractive at small ones. This type of pair interactions is known to lead to the formation of extended stripes structures.\cite{seul,pell,misko} Indeed, such stripes have been observed in theoretical calculations of the vortex matter in thin superconductive films.\cite{cor1,cor3}  

Also, unlike bulk superconductors where the IT superconductivity appears in the interval $-0.41 \tau < \delta \kappa/\kappa_0 < 0.95 \tau$, the crossover regime in thin films is not limited from below, so that a material with any $\kappa$ in the type I domain can be converted into a type II superconductor when the film is thin enough (with the usual limitations on the applicability of the mean field theory). 

The widths $d=10$ and $d=5$ are the most non-trivial because here the bulk IT contributions and those of the stray field are of similar strength both influence the vortex interaction in a large distance interval [Figs. \ref{Fig31} b),c),f),g)]. In this case the interaction becomes even more complicated and can reveal the long- and short-range repulsion and the attraction at intermediate ranges.

\subsection{Many-vortex interactions}

Results of the interaction potential calculated for different many-vortex clusters for the film thicknesses $d=3, 5, 10, 100$~[in the units of $\xi(0)$] and for the GL parameter $\delta \kappa = 0.2, 0, -0.05, -0.11, -0.2, -0.5$ are shown in Fig.~\ref{fig88}. Previous analysis of vortex interactions in clusters in bulk samples\cite{wolf2} has revealed that at $\delta \kappa <0$, i.e., where a single vortex is unstable, the contribution of the many-vortex interactions grows and can stabilize large vortex clusters. This fact can be seen from the results for $d=100$ and $\delta \kappa = -0.11$, where the vortex interaction potential for clusters T and L (see Fig.~\ref{fig88}) of three vortices is fully attractive pointing to instability of these configurations. At the same time, the interaction potentials for S (squares) and P (pentagons) clusters with 4 and 5 vortices, respectively, have shallow minima at finite inter-vortex distances, which stabilizes the clusters. 

As for the two-vortex interactions the influence of the stray fields in negligible for $d=100$ and the results coincide with those for the bulk case. In the other limit of ultra-thin films with $d=3$ the stray field dominates and overcomes the bulk EGL corrections to the vortex interaction. In this limit the vortices in all cluster types are either fully repulsive or attractive at small distances (when the material is in the deep type I regime). The results in Fig. \ref{fig88} for larger $d$ demonstrate an increased role of the many-vortex interactions, which can stabilize larger vortex clusters.  

For example, comparing results for films with $d=100$ and $d=10$, one sees that the stray field determines which of the structures is most stable energetically. In particular, for the film with $d=100$ a triangular T cluster is the most stable [the upper panels of Fig. \ref{fig88} at $\delta\kappa=0.2,0.0$, and $-0.05$]. However, for $d=10$ the linear L cluster becomes most stable, so the stray field makes the line of three vortices a more stable configuration. 

Similarly, for 4-vortex clusters the stray field also alters the most stable configuration: for $d=100$ it is the square S cluster whereas for $d=10$ the triangular T configuration is more preferable. Finally, the pentagon P cluster is most stable among the 5-vortex clusters for both $d=100$ and $d=10$. Yet, the energy difference between the two shapes is larger at $d=10$ - the stray field enhances the stability of P over T clusters. It is possible to expect that the influence of the stray field favours more extended vortex configurations, thereby, leading to formation of vortex chains/stripes. This gives a solid explanation for the observation of the dominating role of the vortex chains in the IT patterns calculated previously for superconducting films.~\cite{cor1,cor3}

\section{Summary} 

This work investigates interactions between vortices in thin superconducting films in the crossover IT regime between types I and II. A particular focus of the study is the interplay between the stray magnetic field and the contributions that lead to a finite IT domain in bulk superconductors. The analysis is done by deriving the formalism that combines the EGL approach, that keeps the leading order corrections to the GL theory, with the stray field contributions to the free energy. The approach thus  takes into account both leading mechanisms that remove the degeneracy of the B point and create a finite IT domain in bulk and film samples. The analysis demonstrates that the degeneracy removal induced by stray fields differs from that in bulk samples. In particular, it changes the profile of the vortex interactions at long ranges. The stray field contribution also has a different temperature dependence and becomes dominant close to the superconductivity transition temperature. 

The derived formalism is applied for the studies of the vortex interactions in various cluster configurations chosen to highlight the different contributions to the vortex interactions in the IT regime. It is shown that the two contributions to the regime of the type crossover are of equal importance when the film thickness is in the interval for which exotic flux configurations appear, as reported previously. The work demonstrates that the stray field contribution changes the most favourable vortex structures. The results suggest that stray field tends to favour extended vortex configurations such as chains of vortices, which explains the spontaneous patterns calculated previously for for IT superconducting films.~\cite{cor1,cor3}    
 
 \acknowledgements
 
 A.V. acknowledges the support from the Brazilian CNPq (Grant No. 307552/2012-8) and the Russian Science Foundation (Project 18-12-00429), used to study non-locality effects in vortex matter. V.S.S. acknowledges the support of the Russian Science Foundation (Project No. 18-72-10118).

\end{document}